# The Volume Rule in the Random Packing Ratio

Honghai Liu[a], Enyong Jiang[b], Chang Q Sun[c] and Bruce Z. Gao[a,*]

[a] Department of Bioengineering, Clemson University, SC 29634, USA.
[b] Institute of Advanced Materials Physics, Faculty of Science, Tianjin University, Tianjin 300072, China.
[c] School of Electrical and Electronic Engineering, Nanyang Technological University, Singapore 639798, Singapore.

**Abstract**

The study on the relationship between the spheres and voids in packing system suggests that the edge effect at the interface between the container and the particles is an important factor lowering the packing ratio. To pack spheres in a container with high packing ratio, an optimized sphere size and an optimized sequence of sphere sizes exist for the packing of single-sized and multi-sized spheres, respectively. We suggest that the concepts of volume and contact should be clearly defined for the packing problem in specific scale.

---

[*] Corresponding author.
E-mail address:
Bruce Z. Gao: zgao@exchange.clemson.edu
Honghai Liu: physhhliu@gmail.com
Chang Q. Sun: ECQSun@ntu.edu.sg

As a classical scientific and mathematical topic, the packing problem of sized particles remains attracting and challenging [1,2]. For spheres of the same size, packing of the face-centered cubic (FCC) lattice has been proved to have the highest possible packing ratio ($\varphi = \pi/\sqrt{18} \approx 0.74$) [3]. For non-spherical particles, the shape may affect the packing ratio [4]. Based on the concept of "coordination numbers" of the particles, Song et al. [5] used statistical theory to construct a phase diagram that provided a unified view of the hard-sphere packing problem, which revealed the effect of the geometric relationship among particles on the random packing. More recently, a "local cell" including the particle and its nearby free space among the system was defined [6] to describe the random packing. By exploring the relationship between the objects with the free space around them, we propose that the interface between the internal wall of the container and the spheres filled is one reason that lowers the packing ratio. The volume concept should be clearly defined when the packing problem is studied in different scale. To pack spheres in a container, an optimized sphere size and a sequence of sphere sizes exist for the packing of single-sized and multi-sized spheres, respectively. A packing ratio of $\varphi \approx 0.76$ by random packing of glass spheres of two sizes has been achieved.

During a packing process, voids form among just-packed neighboring particles. We regard these particles and the voids among them as a subsystem. If the voids can be filled by additional packing particles, the packing ratio will increase. Therefore, the relationship between the particles that remain to be packed and the voids formed by the packed particles is critical. However, few studies have addressed this relationship directly [6]. Man et al. [7] filled voids with fluid to obtain the volume of the voids inside the packing system; this direct method for measuring the packing ratio of a random packing system implied the significance of both particles and voids. Man's fluid molecules served as probes to examine the voids within the system. A similar method was used in studies of computational biology [8], in which water molecules were used as probes to determine the volume of proteins in solution. The positron annihilation lifetime spectroscopy technique uses the same concept to probe the free volume inside materials [9,10]. Correspondingly, the particles that remain to be packed can be thought of as probes to the voids in the packed subsystem.

We used a 2D computer modeling system to study the relationship between the packing particle as a probe and the packed subsystem including the voids. The model packed subsystem consists of 4 disks of identical size (the diameter was $\sigma$, Figure1A). The distance between the centers of the neighboring disks was $d = 1.2\sigma$. The probe was a disk with a variable volume $V_{probe}$ (e.g., the area of the disk). When there was any overlap between the probe and any of the four disks, the probe was regarded as touching the object.

The 4-disk subsystem was placed into a gridded space $\Omega$ (the size of each grid was $0.02\sigma$) with a total volume of $V_\Omega$, which will be surveyed by the probe. When there is no overlap between the probe and the subsystem, all grids covered by the probe would be marked as free. With the number of free grids $N_{free}$ and the number $N_{all}$ of all the grids in the space $\Omega$, the space occupied by the object could be expressed as $V_{object} = V_\Omega \frac{N_{all} - N_{free}}{N_{all}}$, which is the apparent volume of the object in terms of the probe with a volume of $V_{probe}$.

Figure 2A shows that when the $V_{probe}$ is increased from zero, $V_{object}$ changes within two extremes ($3.14\sigma^2 \leq V_{object} < 4.63\sigma^2$, between the blue and red horizontal lines). Figure 2B shows the inaccessibility of the voids in the subsystem to the probe with various volumes $V_{probe}$. When the $V_{probe}$ is sufficient small, all voids inside the container could be probed; thus, the subsystem could be regarded as four separate disks, corresponding to the minimum on the curve in Figure 2A (blue horizontal line). As the $V_{probe}$ is increased, the volume of the voids inside the subsystem that are accessible to the probes became smaller; that is, relative to the probes, the apparent volume of the subsystem is increased. As the $V_{probe}$ is increased indefinitely, the apparent volume of the subsystem approached the volume sheathed by its external tangent (Figures 1B and 2B), which had the asymptotic limit of $4.63\sigma^2$ (the red horizontal line in Figure 2A).

In terms of the model described above, voids inside a packing system exist because the free space surrounded by the subsystem is inaccessible to additional particles of the same size or larger. To increase the packing ratio, the subsystem voids must be accessible to additional particles, and to make these voids accessible, the size of the additional particles must be smaller than the packed ones that form the subsystem (Figure 2) because the packed particles are too large to fill the voids. The simulation results suggested that to achieve high packing ratio using spheres, multisized particles should be used and the smaller particles should be small enough to access the internal of the voids that form in the larger-particle-packed subsystem.

We have investigated the size of the voids using the most compact packing of four identical spheres (radius = $R$), as shown in Figure 3A. There are two characteristic sizes related to the void that forms in the tetrahedron: The radius of the largest sphere (white sphere in Figure 3B) that could pass through the opening of a void is $r_{fc} = (\frac{2\sqrt{3}}{3} - 1)R \approx 0.15R$, and the radius of the largest sphere (white sphere in Figure 3C) that could be contained by a void is $r_{bc} = (\frac{\sqrt{6}}{2} - 1)R \approx 0.22R$. Thus, to make the void inside the tetrahedron in Figure 3A accessible, the radius of the second-size particle should be smaller than $r_{fc} \approx 0.15R$. If we replace one of the four identical spheres in Figure 3A with a larger sphere, the void associated with this replacement will be larger than that in the tetrahedron before the replacement. This implies that when we fill the void formed in the tetrahedron with particles of a size $r$ that is smaller than the $r_{fc}$, the smallest void among the voids newly formed will be the one formed by the filled particles of size $r$ exclusively because other voids will be formed by both the filled particles ($r$) and the packed particles ($R$), which are larger than that formed totally by the filled particles ($r$). Consequently, the multisized spheres for achieving high packing ratio should not be selected randomly; instead, spheres with radii $R_i$ ($i = 1, 2, 3, ...$) that form a sequence of $R_{i+1} \leq 0.15 R_i$ should be selected and used to fill the voids formed by the packed spheres.

To further explore the range of sizes in the sphere sequence, we examined the difference in using large or smaller spheres in the random packing process through another simulation: A large sphere was substituted for FCC-packed smaller spheres that had been placed in a large space (the smaller spheres not replaced by the large sphere remain after placement of the large sphere as shown in Figure 3D, denoted by white color). The total volume of all the smaller spheres that were enclosed or touched (red color in Figure 3D) by the large sphere was denoted as $\sum V_s$ and the volume of the large sphere was $V_l$. The ratio between $\sum V_s$ and $V_l$ indicates which (multiple smaller spheres or one large sphere) will be more efficient in a multisized packing system. The simulation results were plotted in Figure 3E. The ratio between $\sum V_s$ and $V_l$ changed when the size ratio between the larger and smaller spheres is increased ($R_l$ and $R_s$ were the radii of the large and smaller sphere, respectively). After a transitional region (grey area in inset I of Figure 3E), when $R_l / R_s$ was larger than ~11.2, $V_l$ became larger than $\sum V_s$, which means that the large sphere began to be more effective in filling space than the smaller ones when $R_l / R_s > 11.2$. The black horizontal line in Figure 3E represents $\sum V_s / V_l = 1$, and the red horizontal line represents the maximum packing ratio ($\varphi_s \approx 0.74$) of the FCC-packed spheres. We noticed that in Figure 3E, the value of $\sum V_s / V_l$ vibrates and the amplitude of the vibration decreases when $R_l/R_s$ increases, implies that the contacting between the small spheres and the large sphere at the surface of the large sphere also affects the effectiveness that the large sphere filling the space when the small spheres were replaced (edge effect). The results in Figure 3E suggest that 1) if spheres of only one size are used to fill an irregularly shaped space, like the voids, in general trend, the smaller the sphere relative to the space, the higher the packing ratio will be. With the relative size of the packing spheres is decreased, the contacting points between the spheres and the internal wall of the container increases, which may decrease the edge effect and lead to the packing ratio approaching its upper limit of 0.74. In a real packing system, a container must exist. If the a void among the packing system is considered as a container, the interface between the container and the spheres to be filled will be an important factor that affects the packing ratio, smaller spheres may reduce the edge-effect-induced influence on packing ratio; 2) if a large sphere is placed into the packing system with those smaller spheres, when the large sphere's size is 11.2 times larger than that of the smaller spheres, the large sphere will be more effective than those smaller ones in filling the space and thus the packing ratio can be higher than 0.74. Thus, the larger the large sphere, the higher the packing ratio will be. Further more, if the packing ratio inside a single sphere is considered as 100% ("solid" fact), the results also suggest that, to fill a container of specific shape with single-size spheres, an optimized size of the spheres should exist to make good balance between reducing edge effect and taking the advantage of the "solid" fact of single spheres; to make high space using efficiency in a container with filling multi-sized spheres, smaller spheres should be packed to contact the container wall to reduce the edge effect, while larger spheres be packed at the center of the container to take the advantage of the "solid" fact of single spheres. Based on the above discussion, a sequence of spheres with $R_{i+1} = 0.15R_i$ (i = 1, 2, 3, ...) should be used successively to achieve the highest packing ratio: the largest spheres could be filled into the container should be packed before the smaller ones, and after the larger spheres have been packed, the spheres used to fill the subsystem voids should also be the largest among those spheres that are smaller than the size of the voids formed by the packed particles.

To demonstrate this packing concept, we packed glass spheres of two sizes. In the experiments, the radii of the large and small glass spheres were 5 mm and 0.5 mm, respectively. We measured the volume of the spheres and the beaker with water accurately, and the experiments were repeated three times. First, we place the small and large spheres separately into two identical beakers (the volume of the beaker is 274.9 ml) randomly (Figure 4A, 4B), and measured the total volume of the small and large spheres with water. The packing ratios for the small and large spheres in the beakers were $\varphi_s \approx 0.590 \pm 0.003$ and $\varphi_l \approx 0.572 \pm 0.005$, respectively. $\varphi_s > \varphi_l$ indicates that small spheres may decrease the edge effect at the internal surface of the beaker. We, then, carefully added the smaller spheres into the beaker filled with large spheres to form a two size-particle packing system (Figure 4C), without shaking the beaker or stirring the spheres. After packing, we separated and collected the larger and the smaller spheres and measured the total volume of the mixture. The packing ratio of such a system was $\varphi \approx 0.760 \pm 0.002$, higher than the highest packing ratio of single-size spheres $\varphi \approx 0.74$ [3], in accordance with Maxime's results [6].

According to the above discussion of using a sequent of particles with size satisfied $R_{i+1} \approx 0.15 R_i$, we can even achieve a packing ratio close to 1, by successively using particles of sizes with increasing index (*i*) in the sequence. However, problems will emerge while the particle is becoming smaller. There was an basic postulation in the above discussion that each particle is a "solid" sphere, which means that no voids should exist inside the particle, when the particle size used in the series close to the size of atom or even smaller, the basic postulation is broken, and we now meet a paradox: since the size of the smallest particle is comparable to atoms, of which the larger particles are composed, the free space among the atoms inside the larger particles then can not be ignored any longer. According to the rule discussed above, used for the calculation of the volume of the system, the apparent volume of the system becomes smaller, while the size of smaller particles is comparable to atoms. A simple way to avoid the emergence of such a paradox is to set a size limit in the packing system, voids smaller than the size limit will be ignored. The size limit must be smaller than the smallest particle to be packed, in other words, such a small void should never be filled/touched in the packing process.

To explore the reason of the paradox, we need to survey the concept of volume. The critical factor in considering the volume of an object is the definition of its boundary or the "contact" between the object with others. Normally, two objects are considered contact each other, when one stiff object is obstructing the path of the movement of the other stiff object and which, as a result, can not move along its original direction any more, as we have defined the "contact" between probe and the disks in the aforementioned 2D computer modeling system; or when one soft object is distorted through the interaction with others. However, the movements of objects in micro-scale, atom or electron, i.e., are dominated by quantum laws, the boundary of one object or "contact" between objects has to be defined apparently by given rules, as the calculation of molecule volume in previous work (*8*).

In the present research, we studied the possibility of increasing packing ratio of spheres based on the relationship between voids and spheres. The relationship between the container and the spheres is an important factor that affects the packing ratio. To pack spheres with higher packing ratio in a specific container, different rules should be applied according whether single-sized or multi-sized spheres will be packed. The "volume" or "contact" concept of the packing objects should be defined and verified when the pre-enacted packing rule is applied in different scale. The results from the current research may be applied to the study on using resources efficiently in an organization, computer programs and merchandise storage, with the concepts of volume, contact, particles, voids and container being generalized and transformed for specified fields.

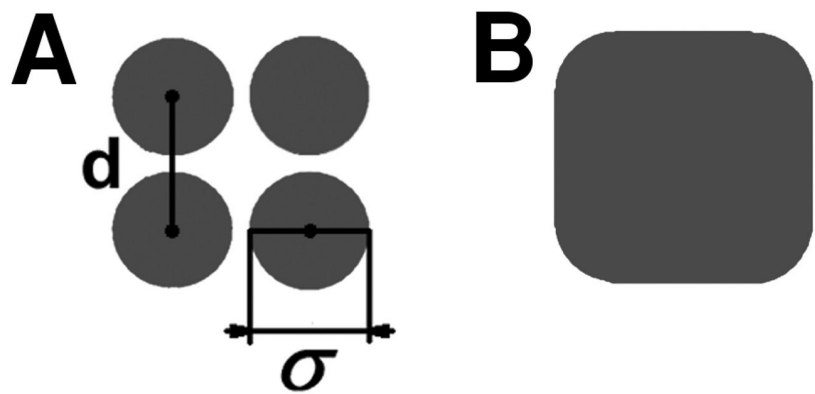

**Figure 1**. (**A**) The modeled subsystem composed of four discs with $d = 1.2\sigma$. (**B**) The space sheathed by the external tangent of the subsystem in (A).

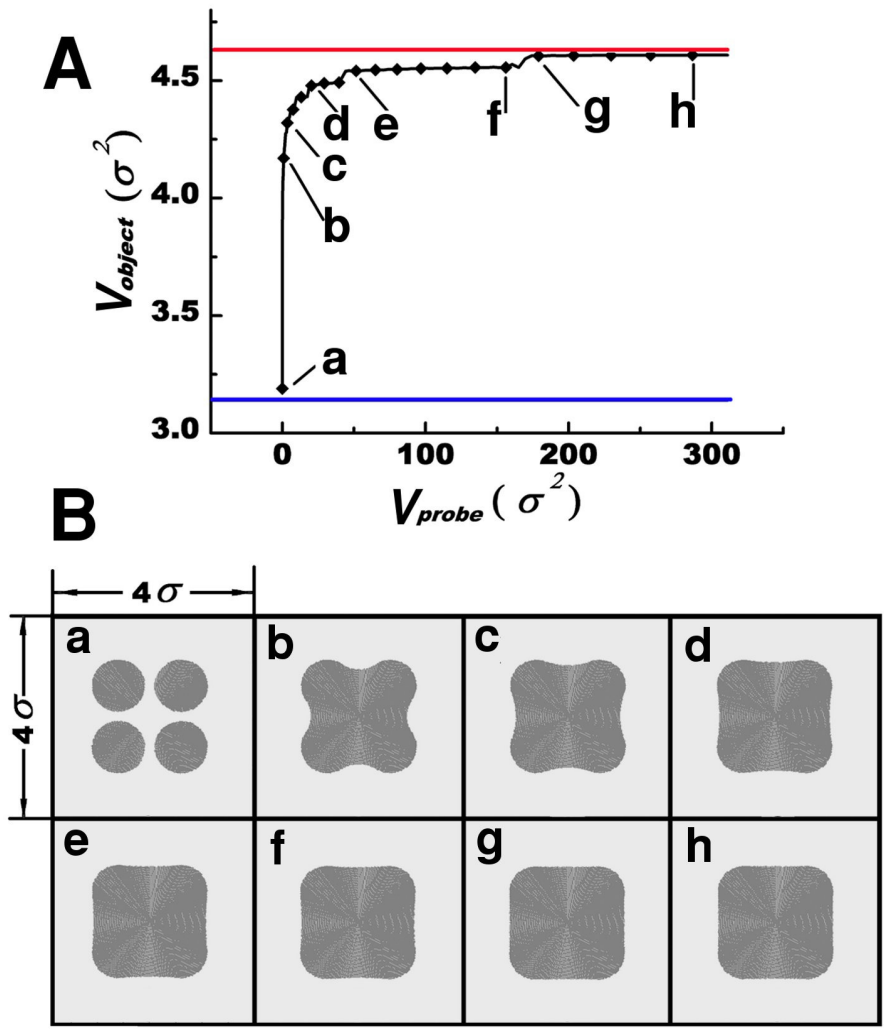

**Figure 2 (A)** The apparent volume of the subsystem vs. the size of the probing particle (beaded black line). The blue horizontal line (the line beneath the curve) denotes the summation of the volume of the four disks ($3.14\sigma^2$); the red horizontal line (the line above the curve) denotes the volume of the sheathed space in Figure 1B ($4.63\sigma^2$). **(B)** The apparent shapes of the subsystem: The shadowed shape in each frame is the apparent shape of the subsystem corresponds to a probing particle with the volume indicated in (A), (a-h).

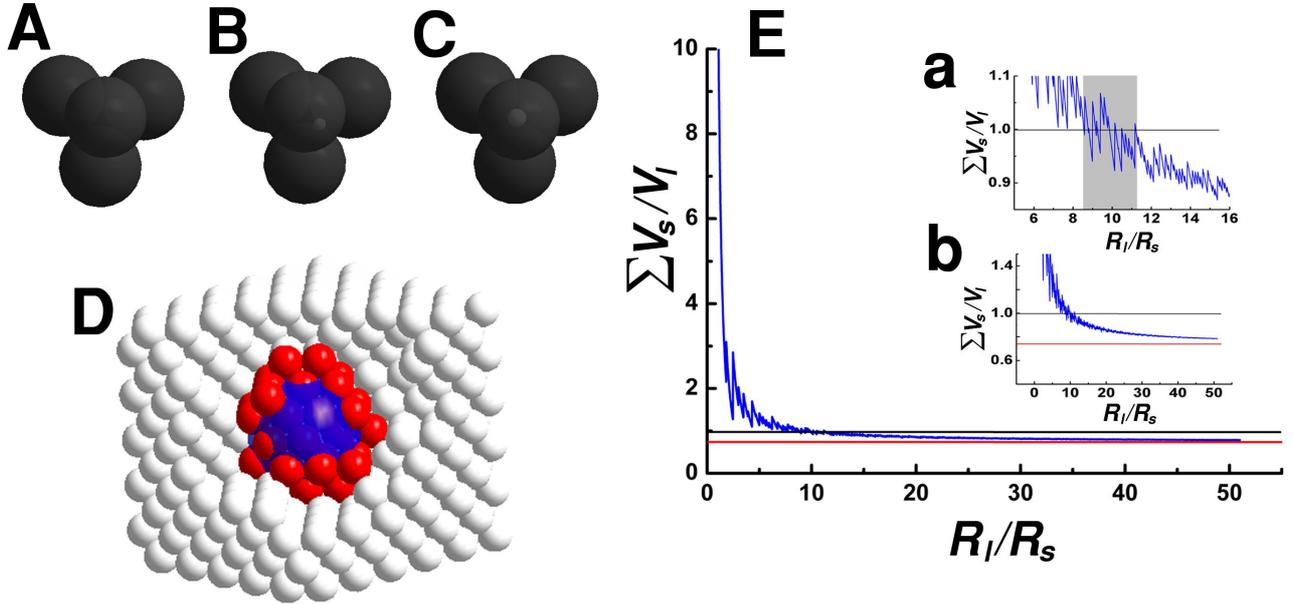

**Figure 3** **(A)** The most compacted packing of four spheres of identical size. **(B)** The largest sphere (white sphere inside the black tetrahedron) that can pass through the surface opening of the tetrahedron. **(C)** The largest sphere (white sphere inside the black tetrahedron) that the void can contain inside the tetrahedron. **(D)** Replacing the FCC-packed small spheres (white spheres) with a large sphere (blue, at the center). All the small spheres enclosed or touched by the large sphere are red. **(E)** The volume ratio between the FCC-packed smaller spheres and the single large sphere used to replace the packed smaller ones. The volume ratio $\sum V_s / V_l$ is plotted as a function of the size ratio $R_l / R_s$ between the larger and smaller spheres. (a) shows the transitional region (grey area) when $\sum V_s / V_l$ changes from value of greater than unity to smaller than unity. (b) shows $\sum V_s / V_l$ approaching the limit of 0.74 when $R_l / R_s$ increases.

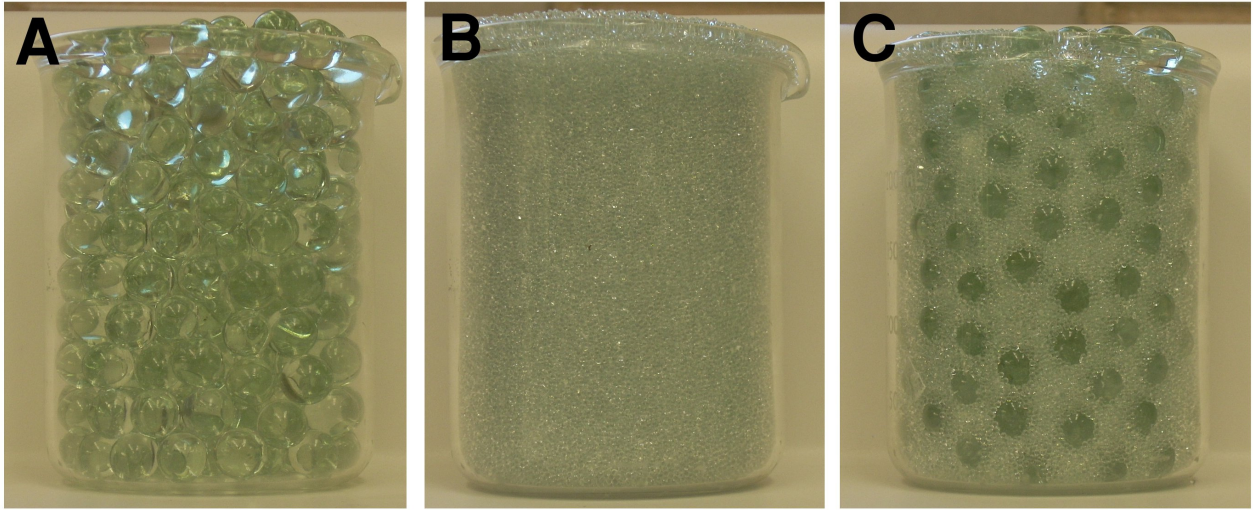

**Figure 4** Random packing of glass spheres. **(A)** Random packing of large spheres (radius = 5 mm). **(B)** Random packing of small spheres (radius = 0.5 mm). **(C)** Random addition of smaller spheres into the packed large spheres in (A).